\documentclass[aps,prl,reprint,superscriptaddress,footinbib,longbibliography]{revtex4-2}

\usepackage[T1]{fontenc}
\usepackage[utf8]{inputenc}
\usepackage{amsmath,amssymb,bm}
\usepackage{graphicx}
\usepackage{xcolor}
\usepackage{tikz}
\usepackage{comment}
\usetikzlibrary{arrows.meta,positioning,calc}
\usepackage[colorlinks=true,citecolor=blue,linkcolor=blue,urlcolor=blue]{hyperref}

\newcommand{\E}{\mathcal{E}}
\newcommand{\D}{\mathcal{D}}
\newcommand{\Coh}{\mathcal{C}}
\newcommand{\Ent}{\mathcal{S}}
\newcommand{\AC}{A_{\Coh}}
\newcommand{\Real}{\operatorname{Re}}
\newcommand{\Imag}{\operatorname{Im}}
\newcommand{\Tr}{\operatorname{Tr}}

\begin{document}

\title{Anomaly Fluctuation Theorem for Quantum Coherence Dynamics}

\author{Kun Zhang}
\email{kunzhang@nwu.edu.cn}
\affiliation{School of Physics, Northwest University, Xi'an 710127, China}
\affiliation{Shaanxi Key Laboratory for Theoretical Physics Frontiers, Xi'an 710127, China}
\affiliation{Peng Huanwu Center for Fundamental Theory, Xi'an 710127, China}
\affiliation{Fundamental Discipline Research Center for Quantum Science and Technology of Shaanxi Province, Xi'an 710127, China}

\author{Hai-Long Shi}
\email{hailong.shi@ino.cnr.it}
\affiliation{INO-CNR and LENS, Largo Enrico Fermi 2, 50125 Firenze, Italy}

\author{Xiao-Hui Wang}
\email{xhwang@nwu.edu.cn}
\affiliation{School of Physics, Northwest University, Xi'an 710127, China}
\affiliation{Shaanxi Key Laboratory for Theoretical Physics Frontiers, Xi'an 710127, China}
\affiliation{Peng Huanwu Center for Fundamental Theory, Xi'an 710127, China}
\affiliation{Fundamental Discipline Research Center for Quantum Science and Technology of Shaanxi Province, Xi'an 710127, China}

\author{Wen-Li Yang}
\email{wlyang@nwu.edu.cn}
\affiliation{Institute of Modern Physics, Northwest University, Xi'an 710127, China}
\affiliation{Shaanxi Key Laboratory for Theoretical Physics Frontiers, Xi'an 710127, China}
\affiliation{Peng Huanwu Center for Fundamental Theory, Xi'an 710127, China}
\affiliation{Fundamental Discipline Research Center for Quantum Science and Technology of Shaanxi Province, Xi'an 710127, China}

\date{\today}

\begin{abstract}
Quantum coherence is a central resource in quantum information science, yet general frameworks and constraints governing its dynamics remain limited. Although quantum coherence lacks a general monotonicity law, we establish an exact integral fluctuation theorem (FT) for coherence dynamics, formulated in terms of Kirkwood-Dirac (KD) quasiprobability trajectories and valid for arbitrary initial states and dynamics. This integral FT deviates from the standard unit-valued form, and the complex deviation, termed the anomaly, quantifies a weighted overlap between the residual final-state coherence and the coherence generated from the dephased input. The real part of the anomaly yields bounds on coherence change, while its imaginary part constrains the second-order moment of the stochastic coherence change weighted by the imaginary parts of the KD quasiprobabilities. Our results establish general statistical constraints on coherence dynamics, demonstrating the utility and broad applicability of FTs for studying quantum-resource dynamics.
\end{abstract}

\maketitle
\emph{Introduction---}Quantum coherence is a fundamental signature of nonclassicality and a central resource for quantum information processing \cite{baumgratz2014,streltsov2017}. Previous work has established important results on coherence evolution in structured settings, specific noise models, and particular state families \cite{lostaglio2015prx,bromley2015,yadin2016,chitambar2016}. However, a unified framework that accommodates arbitrary initial states and dynamics and reveals general constraints on coherence evolution remains lacking. 

Quantum coherence can be quantified in terms of the von Neumann entropy \cite{baumgratz2014}, which is also central to quantum thermodynamics \cite{landiIrreversibleEntropyProduction2021}. Entropy production and its fluctuations obey universal statistical constraints expressed by fluctuation theorems (FTs) \cite{Evans1993ProbabilityOS,jarzynski1997,crooks1999,Evans2002,Esposito2008NonequilibriumFF,campisi2011}. By combining initial-ensemble statistics with trajectory probabilities determined by the dynamics, FTs place initial conditions and dynamical evolution within a single statistical framework, making them powerful tools for studying nonequilibrium dynamics. Research on the interplay between thermodynamics and quantum information has proceeded along two broad directions. In the first, information enters thermodynamic FTs as a correction term \cite{Kim2007FluctuationTF,Sagawa2009GeneralizedJE,Ponmurugan2010GeneralizedDF,Horowitz2010NonequilibriumDF,Sagawa2011NonequilibriumTO,Lahiri2011FluctuationTI,Sagawa2012FluctuationTW,Kundu2012NonequilibriumFT,jevtic2015exchange,MorrisAdesso2018,Park2020InformationFT,Zeng2021NewFT}. In the second, FTs directly describe dynamical changes of quantum-information quantities \cite{Halpern2016JarzynskilikeEF,Halpern2017QuasiprobabilityBT,Micadei2019QuantumFT,Zhang2021ConditionalEP,Zhang2022QuasiprobabilityFT,zhang2026multipartite}.

Information FTs are most naturally formulated when the information quantity obeys a general monotonicity law, analogous to the second law of thermodynamics. For mutual information, the data-processing inequality imposes such a constraint under local quantum channels \cite{schumacherQuantumDataProcessing1996,Wilde2011FromCT}, providing a second-law-like foundation for information FTs \cite{Zhang2022QuasiprobabilityFT,zhang2026multipartite}. Coherence is qualitatively different. For arbitrary initial states and dynamics, its change can have either sign, and no universal monotonicity inequality applies. This raises the fundamental question of whether a nontrivial FT can describe coherence dynamics when coherence itself has no universal direction of change.

In this Letter, we establish an exact FT for the coherence dynamics of arbitrary initial states evolving under general quantum channels. Following the spirit of stochastic thermodynamics, we consider a stochastic coherence change on Kirkwood-Dirac (KD) quasiprobability trajectories. Its KD average equals the standard coherence change, whereas its exponential moment obeys an FT shifted by a complex anomaly \(\AC\). The anomaly quantifies the overlap between residual final coherence and the coherence generated from the dephased input. We prove that maximally incoherent operations, a standard class of free operations in the resource theory of coherence \cite{streltsov2017,chitambar2016}, obey an anomaly-free FT. The real part of the FT yields coherence bounds in which the anomaly acts as a budget constraining coherence growth. Separately, the imaginary part of \(\AC\) constrains the second-order moment of the stochastic coherence change. We illustrate the anomaly FT and its coherence-transfer interpretation analytically in the Jaynes-Cummings model, and test the bounds numerically for randomly sampled qubit states and quantum channels.

\emph{Anomaly FT for coherence dynamics---}We consider a $d$-dimensional system evolving under a completely positive trace-preserving (CPTP) map \(\E\), with input state \(\rho\) and output \(\rho'=\E(\rho)\). We impose no further restrictions on the initial state \(\rho\) or the dynamics \(\E\). Coherence is basis dependent. We fix a reference basis with rank-one projectors \(\{\Pi_x=|x\rangle\langle x|\}\) \footnote{For simplicity, we take the reference bases for the initial and final states to be the same, as is natural in many applications. The following results do not rely on this simplification.}. The coherence can be quantified by \(\Coh(\rho)=\Ent(\D\rho)-\Ent(\rho)\), where \(\Ent(\rho)=-\Tr(\rho\ln\rho)\) is the von Neumann entropy and \(\D(\rho)=\sum_x \Pi_x\rho\Pi_x\) is the dephasing map \cite{baumgratz2014,streltsov2017}. This measure equals the asymptotic distillable coherence under incoherent operations \cite{Winter2016Operational}. The coherence change induced by the dynamics, \(\Delta\Coh=\Coh(\rho')-\Coh(\rho)\), is therefore
\begin{equation}
\Delta\Coh=\Ent(\D\rho')-\Ent(\rho')-\left(\Ent(\D\rho)-\Ent(\rho)\right),
\label{eq:coherence-change}
\end{equation}
which is the central quantity of interest.

In stochastic thermodynamics, the ensemble entropy is associated with the stochastic entropy \(-\ln p\) \cite{seifert2005}. Following this analogy, we introduce the stochastic coherence change
\begin{equation}
\Delta c[\gamma] =
\left(\ln p'_{n'}-\ln p'_{x'}\right)-\left(\ln p_n-\ln p_x\right),
\label{eq:stochastic-coherence-change}
\end{equation}
for trajectories \(\gamma=(n,x,n',x')\). All logarithms and ratios are understood on the active support of the corresponding trajectory weights. Here, \(p_n\) and \(p'_{n'}\) are the eigenvalues of the initial and final states, respectively, defined by
\(\rho=\sum_n p_n\Pi_n\) and \(\rho'=\sum_{n'}p'_{n'}\Pi'_{n'}\), while \(p_x=\Tr(\Pi_x\rho)\) and \(p'_{x'}=\Tr(\Pi'_{x'}\rho')\) are the corresponding populations in the reference basis. The four terms in \(\Delta c\) correspond one-to-one to those in the standard coherence change \(\Delta\Coh\) in Eq.~\eqref{eq:coherence-change}.

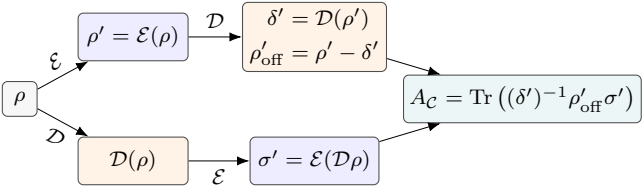
\begin{figure}[t]
\centering
\begin{tikzpicture}[
  font=\footnotesize,
  >=Latex,
  box/.style={draw=black!70, rounded corners=2pt, align=center,
    inner xsep=2.2pt, inner ysep=2pt, minimum height=6.3mm},
  init/.style={box, fill=black!3, inner xsep=1pt, inner ysep=1.5pt,
    minimum width=4.6mm, minimum height=4.6mm},
  midbox/.style={box, text width=1.55cm},
  actual/.style={midbox, fill=blue!7, text width=1.3cm},
  dephased/.style={midbox, fill=orange!10, text width=1.3cm},
  core/.style={midbox, fill=orange!10, text width=1.75cm},
  reference/.style={midbox, fill=blue!7},
  anomaly/.style={box, fill=teal!7, text width=2.99cm},
  arrow/.style={->, line width=0.38pt}
]
\node[init] (rho) at (0,0) {\(\rho\)};
\node[actual] (rhop) at (1.5,0.82) {\(\rho'=\E(\rho)\)};
\node[dephased] (drho) at (1.5,-0.82) {\(\D(\rho)\)};
\node[core] (core) at (3.9,0.82)
  {\(\begin{aligned}
  \delta'&=\D(\rho')\\[-1pt]
  \rho'_{\mathrm{off}}&=\rho'-\delta'
  \end{aligned}\)};
\node[reference] (sigma) at (3.9,-0.82)
  {\(\sigma'=\E(\D\rho)\)};
\node[anomaly] (ac) at (6.65,0)
  {\( \AC=\Tr\left((\delta')^{-1}\rho'_{\mathrm{off}}\sigma'\right)\)};

\draw[arrow] (rho) -- node[above,sloped] {\(\E\)} (rhop);
\draw[arrow] (rho) -- node[below,sloped] {\(\D\)} (drho);
\draw[arrow] (rhop) -- node[above] {\(\D\)} (core);
\draw[arrow] (drho) -- node[below] {\(\E\)} (sigma);
\draw[arrow] (core) -- (ac);
\draw[arrow] (sigma) -- (ac);
\end{tikzpicture}
\caption{Schematic illustration of the anomaly \(\AC\) in the coherence FT, Eq.~\eqref{eq:coherence-ft}. Here \(\rho\) and \(\rho'=\mathcal E(\rho)\) are the initial and final states, while \(\sigma'=\mathcal E(\mathcal D\rho)\) is the output produced by the CPTP map \(\mathcal E\) from the dephased input \(\mathcal D(\rho)\). The final state \(\rho'\) is decomposed into the dephased part \(\delta'=\mathcal D(\rho')\) and the residual coherence \(\rho'_{\mathrm{off}}=\rho'-\delta'\).}
\label{fig:mechanism}
\end{figure}

To relate a stochastic quantity to its ensemble average, one must also assign weights to the underlying trajectories. Unlike classical stochastic dynamics, however, a quantum process does not, in general, admit a classical path-probability description. More formally, the statistics of quantum evolution are not always representable by a classical stochastic process, as reflected, for example, in violations of the Leggett-Garg inequality \cite{leggett1985,emary2014}. We therefore represent the quantum trajectories by the KD quasiprobability \cite{Perarnau2017,Lostaglio2018,lostaglio2023kd,arvidsson2024review,GherardiniDeChiara2024}
\begin{equation}
\mathcal{Q}[\gamma]=
\Tr\!\left(\Pi'_{x'}\Pi'_{n'}\,\E\!\left(\Pi_x\Pi_n\rho\right)\right).
\label{eq:kd-trajectory}
\end{equation}
The KD quasiprobability is normalized but need not be positive or real. It reproduces the average coherence change,
$\langle \Delta c\rangle_{\mathcal Q}=\Delta\Coh$, where we use the convention \(\langle f\rangle_{\mathcal Q} = \sum_\gamma \mathcal Q[\gamma]f[\gamma]\). For a generic initial state and CPTP map, the coherence change \(\Delta\Coh\) can have either sign. The stochastic coherence change nevertheless obeys an exact integral FT.

\emph{Theorem 1---Denote \(\delta'=\D(\rho')\), \(\sigma'=\E(\D\rho)\), and \(\rho'_{\mathrm{off}}=\rho'-\delta'\). Assume that $\delta'$ is full rank. The stochastic coherence change \(\Delta c\) satisfies
\begin{equation}
\left\langle e^{\Delta c}\right\rangle_{\mathcal Q} = 1+\AC,\qquad \AC=\Tr\!\left((\delta')^{-1}\rho'_{\mathrm{off}}\sigma'\right),
\label{eq:coherence-ft}
\end{equation}
where $\AC$ is termed the FT anomaly.}

The proof is provided in the Supplementary Material (SM) \cite{SM}. If the anomaly \(\AC\) vanishes, then the stochastic coherence change satisfies a standard integral FT. Figure~\ref{fig:mechanism} provides a schematic illustration of the anomaly \(\AC\). The three operators in \(\AC\) have clear physical meanings: \(\delta'\) is the dephased final state, \(\rho'_{\mathrm{off}}\) is the off-diagonal part of the final state in the reference basis, which represents the final coherence, and \(\sigma'\) is the reference output generated from the dephased input $\mathcal D(\rho)$. The complex anomaly \(\AC\) quantifies the phase-sensitive overlap between the residual final coherence \(\rho'_{\mathrm{off}}\) and the coherence generated from the dephased input \(\sigma'\). Its magnitude measures the strength of this overlap, while the sign of its real part distinguishes constructive or destructive interference. Note that the anomaly is determined solely by the outputs of \(\mathcal E\) for two different inputs. 

If \(\delta'\) is not full rank and the support of \(\sigma'\) is not contained within that of \(\delta'\), the anomaly FT requires additional terms accounting for population leakage, analogous to the absolute-irreversibility correction in conventional FTs \cite{Murashita2014,Funo2015,Murashita2017}. We next identify anomaly-free mechanisms that connect our result to existing FTs and clarify the anomaly's resource-theoretic meaning. We then derive anomaly-dependent bounds on coherence dynamics.

\emph{Anomaly-free mechanisms---}The reference state \(\sigma'=\E(\D\rho)\) is generated from an incoherent input. Therefore, the anomaly is constrained by the channel's cohering power, defined as the maximum coherence that the channel can generate from an incoherent state and given by \(\mathcal P(\mathcal E) = \max_{\omega\in \mathcal I_{\D}}\Coh(\E\omega)\). The set of incoherent states is \(\mathcal I_{\D}=\{\omega:\D(\omega)=\omega\}\) \cite{mani2015cohering}. Specifically, the anomaly is bounded by
\begin{equation}
|\AC|\leq
\frac{2\sqrt{\Coh(\rho')\,\mathcal P(\mathcal E)}}{p'_{\min}},
\label{eq:cohering-power-anomaly-bound}
\end{equation}
where \(p'_{\min}=\min_{x'}p'_{x'}\) is the smallest nonzero final population. The derivation is given in the SM \cite{SM}.

The bound \eqref{eq:cohering-power-anomaly-bound} reveals two simple anomaly-free mechanisms. The first is final dephasing, in which the output coherence vanishes, \(\mathcal C(\rho')=0\), and hence \(\AC=0\). The final-dephasing coherence FT was first reported in Ref.~\cite{zhang2026multipartite}. The second is reference incoherence, in which the reference output is diagonal in the final basis, \(\sigma'=\D(\sigma')\). A maximally incoherent CPTP map has zero cohering power, because it maps incoherent states to incoherent states \cite{streltsov2017}. This immediately gives the following corollary.

\emph{Corollary 2---If the channel is maximally incoherent, \(\E(\mathcal I_{\D})\subseteq\mathcal I_{\D}\), then the coherence FT is anomaly-free.}

For a maximally incoherent channel, the relative-entropy coherence is independently known to be nonincreasing as a consequence of resource-theoretic monotonicity. The anomaly-free FT therefore provides a stochastic counterpart to coherence monotonicity. Note that the anomaly contains an inverse-population weighting through \((\delta')^{-1}\sigma'\). The density operator \(\delta'\) gives the final-state coherence, \(\Coh(\rho')=\mathcal S(\delta')-\mathcal S(\rho')\). Replacing \(\delta'\) with \(\sigma'\) removes the anomaly and the anomaly-free FT describes the complex entropy production of the quantum channel \cite{kwon2019}. However, the stochastic average in that formulation is no longer the coherence change \(\Delta\mathcal C\). 

Thermodynamic FTs with anomalies have also been reported recently. The anomaly arises from initial energetic coherence in Ref.~\cite{levy2020} and from noncommuting conserved charges in Ref.~\cite{Upadhyaya2024}. The anomaly in our FT arises from a mismatch between \(\Delta c\) and the KD quasiprobability \(\mathcal Q\), which results in the associated reverse-trajectory quasiprobability distribution being unnormalized. Such a mismatch is unavoidable because \(\Delta c\) and \(\mathcal Q\) are jointly chosen to satisfy the physical requirement \(\langle\Delta c\rangle_{\mathcal Q}=\Delta\mathcal C\). 



%



\emph{Bounds on coherence change---}The standard FT imposes strong constraints on the statistics of stochastic variables \cite{merhavStatisticalPropertiesEntropy2010}. From Jensen's inequality, \(e^{\langle f\rangle}\leq \langle e^f\rangle\), the standard FT \(\langle e^{-f}\rangle=1\) yields the second-law inequality \(\langle f\rangle\geq 0\) for the mean. Quasiprobability distributions, however, can contain negative or complex weights, so this conventional argument does not apply directly. Although Jensen-type inequalities can remain valid for signed weights \cite{Horvath2024,Yu2026Negative}, a general statistical treatment of FTs formulated in terms of complex quasiprobabilities has yet to be established. We address this issue using two complementary strategies and derive two nontrivial bounds on the coherence change.

The first strategy identifies an additional condition under which the bound obtained by combining the FT with Jensen's inequality remains valid even in the presence of negative quasiprobabilities.

\emph{Theorem 3---Let \(d\) denote the dimension of \(\rho\). If \(\mathrm{Re}\,\AC\geq d-2\), then the coherence change is bounded in terms of the real part of the anomaly $\AC$ as
\begin{equation}
\label{eq:coherence-bound}
\Delta\mathcal C \leq \ln(1+\mathrm{Re}\,\AC).
\end{equation}}

The proof is given by combining the data-processing inequality with an operator-logarithm bound. The threshold \(\mathrm{Re}\,\AC\geq d-2\) is sharp. Examples in the SM show that the bound can fail when \(\mathrm{Re}\,\AC<d-2\) \cite{SM}. For a qubit (\(d=2\)), an anomaly-free FT (\(\AC=0\)) precludes strict coherence growth, while a positive real anomaly provides a coherence-growth budget of \(\ln(1+\Real\AC)\). Physically, if the channel generates no coherence from the dephased initial state, then coherence cannot increase. More generally, even if the channel generates coherence from the dephased initial state, coherence may still not increase when the initial coherence interferes destructively with the coherence generated from the populations. In the regime \(\mathrm{Re}\,\AC<d-2\), violations of Eq.~\eqref{eq:coherence-bound} are attributable to negative real KD weights.





\begin{figure}[t]
  \centering
  \includegraphics[width=\columnwidth]{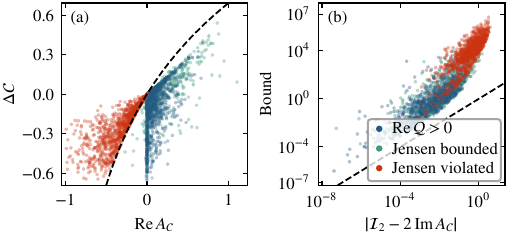}
\caption{Numerical tests of the coherence bounds in Eqs.~\eqref{eq:coherence-bound} and \eqref{eq:imaginary-second-order-bound} for randomly sampled qubit states and CPTP maps. (a) Coherence change \(\Delta\Coh\) versus \(\Real\AC\). The dashed curve marks saturation of Eq.~\eqref{eq:coherence-bound}. Blue points correspond to data with nonnegative real KD weights. Green (red) points correspond to signed real KD weights that satisfy (violate) Jensen's inequality. (b) Left-hand side \(|\mathcal I_2-2\Imag\AC|\) versus the corresponding right-hand side of Eq.~\eqref{eq:imaginary-second-order-bound} on logarithmic axes. The dashed diagonal marks saturation. The data points from (a) and (b) are generated from the same random samples.}
\label{fig:bound}
\end{figure}

We numerically test the coherence-change bound \eqref{eq:coherence-bound} using both randomly sampled qubit states and CPTP maps in Fig.~\ref{fig:bound}(a), which shows \(\Delta\mathcal C\) against \(\Real\AC\), with the dashed curve marking the threshold \(\ln(1+\Real\AC)\). We divide the samples into three classes: blue points have nonnegative real KD weights, green points have signed real KD weights but satisfy the Jensen inequality, and red points violate it. All samples with nonnegative real KD weights lie below the threshold. The red points above the threshold occur only when $\Real\AC<0$ and witness the negativity of the quasiprobability trajectories.


The second strategy separates the negative KD weights from the nonnegative ones, allowing Jensen's inequality to be applied to the latter. This yields a negativity-corrected bound on the coherence change.

\emph{Theorem 4---Let \(\tilde q = \sum_{\Real\mathcal Q<0}|\Real \mathcal Q|\) denote the sum of all negative real KD weights. Suppose that the stochastic coherence change is bounded as \(|\Delta c|\leq r_c\). Then
\begin{equation}
\label{eq:coherence-bound_negative_corrected}
\Delta\mathcal C \leq (1+\tilde q)\ln\left(\frac{1+\Real\AC + \tilde qe^{r_c}}{1+\tilde q}\right) + \tilde q r_c.
\end{equation}}

The proof is given in the SM \cite{SM}. As a consistency check, if all real KD weights are nonnegative, then \(\tilde q=0\), and Eq.~\eqref{eq:coherence-bound_negative_corrected} reduces to the bound \eqref{eq:coherence-bound}. The corrected bound is rigorous but generally not tight. However, unlike the bound in Theorem 3, the corrected bound imposes no threshold condition on \(\Real\AC\). Cases that violate the Jensen-type bound \eqref{eq:coherence-bound} in the regime \(\Real\AC<d-2\) still satisfy the negativity-corrected bound \eqref{eq:coherence-bound_negative_corrected}.

\emph{Imaginary KD weights and the second-order moment---}Although \(\mathcal Q\) is generally complex, the KD mean $\langle \Delta c\rangle_{\mathcal Q}=\Delta\Coh$ is real. Together with the normalization condition $\sum_\gamma \mathcal Q[\gamma] = 1$, this implies \(\sum_\gamma\Imag\mathcal Q[\gamma]=0\) and \(\sum_\gamma\Imag\mathcal Q[\gamma]\Delta c[\gamma]=0\). Therefore, only the second- and higher-order imaginary moments of \(\Delta c\) can be nonzero and contribute to \(\Imag\AC\) in the FT. Let 
\begin{equation}
\mathcal I_2=\sum_\gamma \Imag\mathcal Q[\gamma](\Delta c[\gamma])^2
\end{equation}
denote the second-order moment of the stochastic coherence change weighted by the imaginary part of the KD quasiprobability. The anomaly FT then implies that the second-order contribution to \(\Imag\AC\) is \(\mathcal I_2/2\). More precisely, the following result bounds the higher-order remainder.

\emph{Theorem 5---Let \(q_I=\sum_\gamma|\Imag\mathcal Q[\gamma]|/2\) denote half the total variation of the imaginary KD weights, and suppose that $|\Delta c|\leq r_c$. Then
\begin{equation}
\left|\mathcal I_2-2\Imag\AC\right|\le \frac{2}{3}q_I e^{r_c}r_c^3.
\label{eq:imaginary-second-order-bound}
\end{equation}}

The proof is given in the SM \cite{SM}. For \(r_c\ll1\), Eq.~\eqref{eq:imaginary-second-order-bound} yields \(\mathcal I_2=2\Imag\AC+\mathcal O(q_I r_c^3)\). Thus, up to cubic and higher-order terms, \(\Imag\AC\) directly reflects the contribution of \(\Imag\mathcal Q\) to the second-order moment of the stochastic coherence change. When the stochastic coherence change $\Delta c$ is small, the imaginary anomaly directly measures fluctuations weighted by the imaginary KD weights, even though these weights do not contribute to the average coherence change.

Figure~\ref{fig:bound}(b) tests the imaginary second-order bound on the same randomly generated samples presented in Fig.~\ref{fig:bound}(a). The horizontal coordinate is \(|\mathcal I_2-2\Imag\AC|\), and the vertical coordinate is the corresponding analytic upper bound in Eq.~\eqref{eq:imaginary-second-order-bound}. All points lie on the allowed side of the dashed equality line, verifying the bound for all three classes. The Jensen-inequality-violating samples extend farther into the upper-right region, where both the remainder \(|\mathcal I_2-2\Imag\AC|\) and its upper bound are larger. This trend suggests that Jensen violations are accompanied by stronger imaginary fluctuations. 


\emph{Anomaly FT in the Jaynes-Cummings model---}We illustrate the anomaly FT and its coherence bound in the Jaynes-Cummings (JC) model, a canonical setting for the coherent exchange of excitations between a two-level atom and a single field mode \cite{jaynes1963,shore1993}. Within the rotating-wave approximation and with \(\hbar=1\), the Hamiltonian is \(H_{\mathrm{JC}}=\omega_F a^\dagger a+\omega_A\sigma_+\sigma_-+g(\sigma_+a+\sigma_-a^\dagger)\), where \(\omega_F\) and \(\omega_A\) are the field-mode and atomic transition frequencies. We consider resonant coupling, \(\omega_A=\omega_F\), and parameterize the evolution by the dimensionless interaction angle \(\theta=gt\).

\begin{figure}[t]
  \centering
  \includegraphics[width=\columnwidth]{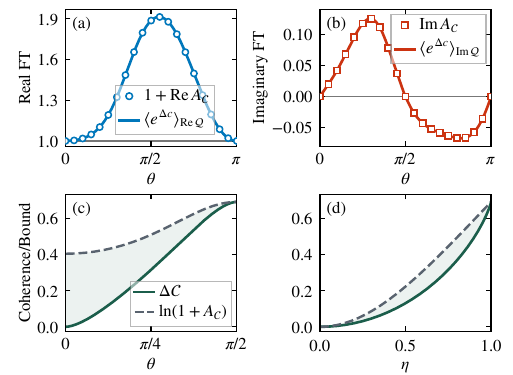}
\caption{Anomaly FT and coherence transfer in the resonant JC model. (a) Real and (b) imaginary parts of the anomaly FT, Eq.~\eqref{eq:coherence-ft}. (c) Transferred coherence \(\Delta\Coh\) and the bound \(\ln(1+\AC)\) versus \(\theta=gt\) for \(\rho_A=|g\rangle\langle g|\), \(\eta=1\), and \(\bar n=1/2\). (d) The same quantities versus the field coherence visibility \(\eta\) at \(\bar n=1/2\) and \(\theta=\pi/2\).}
\label{fig:JC_model}
\end{figure}

We restrict the initial field to the single-rail subspace spanned by the vacuum state \(|0\rangle\) and the one-photon state \(|1\rangle\), and parameterize its initial state as
\begin{equation}
\rho_F(\eta,\bar n)=
\begin{pmatrix}
1-\bar n & \eta\sqrt{\bar n(1-\bar n)}\\
\eta\sqrt{\bar n(1-\bar n)} & \bar n
\end{pmatrix}.
\label{eq:jc-field-state}
\end{equation}
Here \(\bar n=\Tr(\rho_F a^\dagger a)\) is the mean photon number, and \(0\leq\eta\leq1\) is the normalized coherence visibility in the basis $\{|0\rangle,|1\rangle\}$. We choose the convention that the off-diagonal element is real. The limits \(\eta=1\) and \(\eta=0\) describe, respectively, a pure single-rail superposition and its fully dephased mixture with the same photon-number statistics. Figures~\ref{fig:JC_model}(a) and (b) verify the real and imaginary parts of Eq.~\eqref{eq:coherence-ft}, respectively. The nonzero imaginary part of the FT originates from coherent atom-field interference encoded in the complex KD weights.


The anomaly $\AC$ serves as a coherence-transfer budget and thus provides a refined indicator of coherence transfer in transcoherent-state studies \cite{goldberg2020transcoherent,goldberg2023beyondtranscoherent}. We initialize the atom in \(\rho_A=|g\rangle\langle g|\) and measure coherence in the energy basis \(\{|g\rangle,|e\rangle\}\). The atomic input is incoherent in this basis, namely \(\D(\rho_A)=\rho_A\). During the first half Rabi cycle, the excited-state population and atomic coherence amplitude are \(p_e=\bar n\sin^2\theta\) and \(|\rho_{ge}|=\eta\sqrt{\bar n(1-\bar n)}\sin\theta\), respectively. Whenever both output populations are nonzero, the anomaly is
\begin{equation}
\AC
=\eta^2\frac{1-\bar n}{1-\bar n\sin^2\theta}.
\label{eq:jc-anomaly}
\end{equation}
This expression applies for \(0<\bar n<1\), \(0\leq\eta\leq1\), and \(\sin\theta\neq0\). Equivalently, \(\AC=|\rho_{ge}|^2/(p_gp_e)\). Thus, \(\AC\) measures the squared atomic coherence visibility supported by the output populations.


For the ground-state input $\rho_A = |g\rangle\langle g|$, the KD weights are nonnegative, so \(\Delta\Coh\leq\ln(1+\AC)\). For fixed \(0<\bar n<1\) and \(\eta=1\), the anomaly \(\AC\) is maximized at \(\theta=\pi/2\), indicating the maximal coherence-transfer capability. With equal initial field populations $\bar n=1/2$, the coherence transfer reaches its maximum \(\Delta\Coh=\ln2\) with \(\AC=1\), saturating the coherence bound \eqref{eq:coherence-bound}. See Fig.~\ref{fig:JC_model}(c). With the optimal evolution angle $\theta=\pi/2$ and equal initial field populations $\bar n=1/2$ fixed, the anomaly $\AC=\eta^2$ increases monotonically with the field visibility $\eta$, indicating that the coherence-transfer capability decreases as $\eta$ decreases. See Fig.~\ref{fig:JC_model}(d). The gap between \(\Delta\Coh\) and \(\ln(1+\AC)\) is strictly positive for \(0<\eta<1\) and closes at \(\eta=0\) and \(\eta=1\). In particular, the vanishing anomaly $\AC$ at $\eta=0$ directly demonstrates that the JC model with a dephased field input cannot transfer coherence to the atom \cite{SM}.

\emph{Conclusion and outlook---}We have formulated an exact quantum FT for coherence dynamics, Eq.~\eqref{eq:coherence-ft}, valid for general initial states and quantum dynamics. The FT is governed by a complex anomaly \(\AC\), which quantifies a weighted overlap between the final-state coherence and the reference coherence generated from the dephased initial state. From this anomaly FT, we derive two upper bounds on the coherence change: the Jensen-type bound in Eq.~\eqref{eq:coherence-bound} and the negativity-corrected bound in Eq.~\eqref{eq:coherence-bound_negative_corrected}. The real part of the anomaly plays a central role in bounding coherence changes and can serve as a budget for coherence growth. We further show that, to leading order, \(\Imag\AC\) determines the contribution of \(\Imag\mathcal Q\) to the second-order moment of the stochastic coherence change, with the higher-order deviation bounded by Eq.~\eqref{eq:imaginary-second-order-bound}.

More broadly, our results demonstrate that quasiprobability FTs with possible anomalies provide a promising framework for studying quantum-information dynamics. Rather than signaling a breakdown of the FT, the anomaly encodes dynamical information absent from the conventional unit-valued FTs and thereby furnishes additional constraints. This perspective invites extensions to other quantum resources and correlations, including entanglement asymmetry \cite{Ares2022EntanglementAA,Ares2025TheQM} and quantum discord \cite{Ollivier2001QuantumDA,modi2012classical}. A general statistical theory of FTs for quantum-information dynamics must accommodate signed or complex KD trajectory weights, calling for a quasiprobabilistic counterpart of the classical theory that combines probability distributions with FTs \cite{merhavStatisticalPropertiesEntropy2010}. The recently derived quasiprobability thermodynamic uncertainty relation \cite{Yoshimura2026QUTR} provides an important step in this direction. Developing uncertainty relations tailored to quasiprobability FTs with anomalies is therefore a natural direction for future work and may clarify how negativity, imaginary KD weights, and dynamical anomalies jointly constrain fluctuations of quantum resources.

\emph{Acknowledgments.---} This work was supported by the NSFC (No.12305028, No.12275215, No.12247103, and No.12434006), and the Youth Innovation Team of Shaanxi Universities. KZ is supported by the China Postdoctoral Science Foundation under Grant Number 2025M773421, Shaanxi Province Postdoctoral Science Foundation under Grant Number 2025BSHYDZZ017, and Scientific Research Program Funded by Education Department of Shaanxi Provincial Government (Program No. 24JP186). HLS was supported by the Horizon Europe programme HORIZONCL4-2022-QUANTUM-02-SGA via Project No. 101113690 (PASQuanS2.1). 

\makeatletter
\renewcommand{\bibfont}{\small\linespread{0.94}\selectfont\@clubpenalty\clubpenalty}
\immediate\write\@auxout{\string\citation{apsrev42Control}}
\makeatother

%

\end{document}